\documentclass[structabstract]{aph}  
\usepackage{graphicx}
\usepackage{epstopdf}
\usepackage{amsmath}
\usepackage{txfonts}
\usepackage{natbib}
\bibpunct{(}{)}{;}{a}{}{,}
\begin{document}
   \title{Excitation and abundance study of CO$^+$ in the interstellar medium}

   \subtitle{}

   \author{P. St\"auber
%          \inst{1}
          \and
          S. Bruderer}

   \institute{Institute for Astronomy, ETH Zurich,
              8093 Zurich, Switzerland\\
              \email{pascal.staeuber@phys.ethz.ch}}

% \abstract{}{}{}{}{} 
% 5 {} token are mandatory
 
\abstract
{Observations of CO$^+$ suggest column densities on the order $10^{12}$~cm$^{-2}$ that can not be reproduced by many chemical models. CO$^+$ is more likely to be destroyed than excited in collisions 
with hydrogen. An anomalous excitation mechanism may thus have to be considered when interpreting CO$^+$ observations. Other uncertainties in models are the chemical network, the gas temperature or the geometry of the emitting source. Similar is true for other reactive ions that will be observed soon with the {\it Herschel Space Observatory}.}
{Chemical constraints are explored for observable CO$^+$ abundances. The influence of an anomalous excitation mechanism on CO$^+$ line intensities is investigated. Model results are compared to observations.}
{Chemical models are used to perform a parameter study of CO$^+$ abundances. Line fluxes are calculated for $N(\mathrm{CO^+})=10^{12}$~cm$^{-2}$ and different gas densities and temperatures using a non--LTE escape probability method. The chemical formation and destruction rates are considered explicitly in the detailed balance equations of the radiative transfer. In addition, the rotational levels of CO$^+$ are assumed to be excited upon chemical formation according to a {\it formation temperature}. Collisional excitation by atomic and molecular hydrogen as well as by electrons is studied for conditions appropriate to dense photon-dominated regions (PDRs) and star-forming environments.}
{Chemical models are generally able to produce high fractional CO$^+$ abundances ($x({\mathrm{CO^+}}) \approx 10^{-10}$). In a far-ultraviolet (FUV) dominated environment, however, high abundances of CO$^+$ are only produced in regions with a Habing field G$_0 \ga 100$ and $T_{\mathrm{kin}} \ga 600$~K, posing a strong constraint on the gas temperature. For gas densities $\la 10^6$~cm$^{-3}$ and temperatures $\ga 600$~K, the combination of chemical and radiative transfer analysis shows little effect on intensities of CO$^+$ lines with upper levels $N_{\mathrm{up}} \le 3$. Significantly different line fluxes are calculated with an anomalous excitation mechanism, however, for transitions with higher upper levels and densities $\ga 10^6$~cm$^{-3}$. The Herschel Space Observatory is able to reveal such effects in the terahertz wavelength regime. Ideal objects to observe are protoplanetary disks with densities $\ga 10^6$~cm$^{-3}$ . It is finally suggested that the CO$^+$ chemistry may be well understood and that the abundances observed so far can be explained with a  high enough gas temperature and a proper geometry.}
{}


\maketitle
 
%________________________________________________________________

\section{Introduction}
\label{intro}

Since the first detection of the molecular ion CO$^+$ in the interstellar medium by \citet{LaWM93}, subsequent observations have been carried out successfully toward several photodissociation regions (PDRs), reflection nebulae and planetary nebulae \citep{HoJv95,StSS95,FuRG03}, protostellar envelopes \citep{CeCW97,StBJ07} and even galaxies \citep{FuBM00,FuGG06}. As the $N=1\rightarrow 0$ transition is blended by atmospheric O$_2$, CO$^+$ is most commonly observed in the $N=2\rightarrow 1$ and $N=3\rightarrow 2$ rotational transitions. The fine-structure levels $J=N \pm \frac{1}{2}$ of the lowest few rotational transitions are close in frequency and they are thus usually observed simultaneously with modern spectrometers that offer bandpasses of $1$--$2$~GHz. Higher rotational transitions have been observed with ISO toward the low-mass protostar IRAS $16293$--$2422$ \citep{CeCW97}.

The reactive ion is believed to be sensitive to far-ultraviolet (FUV) photons and X-rays and thus to be a tracer for PDRs and X-ray dominated regions (XDRs). Chemical models have confirmed this by showing that the presence of FUV photons or X-rays enhances the CO$^+$ abundance by several orders of magnitude \citep{StDa95,StDv05,SpMe07}. The observed CO$^+$ column densities ($N(\mathrm{CO^+}) \sim 10^{12}$~cm$^{-2}$), however, pose a real challenge to PDR and XDR models. They fail to reproduce these abundances often by more than an order of magnitude. It should be mentioned though, that 'traditional' models tread the chemistry either in a plane-parallel \citep{StDa95} or spherical \citep{StDv05} one dimensional geometry. A careful treatment of the geometry could therefore be a solution to this problem. The geometry of the emitting source is included in the work of \citet{JaSH95} for the Orion Bar, \citet{FuGU08} for the starburst galaxy M~$82$ and \citet{BrDV09} for the massive star-forming region AFGL~$2591$. Their results will be discussed in Sect.~\ref{chemistry}.

An interesting feature of the observed CO$^+$ emission lines is that the rotational excitation temperatures appear to be as low as $T_{\mathrm{ex}} \approx 10$~K \citep{LaWM93,FuRG03}. Considering also the fact that the timescales for rotational excitation and chemical destruction of CO$^+$ are similar, suggests that CO$^+$ is destroyed before its translational motions become thermalized \citep{Black98}. In addition, CO$^+$ might be excited upon formation.  Nascent excitation effects occur when some of the enthalpy change in a reaction goes into rotational (or vibrational) excitation of CO$^+$. On the other hand, if CO$^+$ is formed through ionization of CO by energetic photons or particles, it will tend to inherit the rotational excitation of its parent because the relatively heavy nuclei can not respond during the rapid electronic ionization process.  It is therefore not clear how the rotational CO$^+$ levels are excited and what influence an anomalous excitation mechanism has on the line fluxes.

Motivated by the shortcomings of traditional PDR and XDR models and the considerations in the previous paragraph, the aim of this paper is to study the excitation of CO$^+$ by combining chemical models and radiative transfer analysis. Uncertainties in chemical models are the gas temperature and the chemical network. Thus, we first explore the chemistry of CO$^+$ in a general parameter study to constrain the physical gas conditions required to produce large fractional abundances (Sect.~\ref{chemistry}). We then make use of RADEX, a computer program for fast non--LTE analysis of interstellar line spectra \citep{vdTBS07}, to calculate the rotational excitation of CO$^+$, including the formation and destruction rates explicitly. The method is described in Sect.~\ref{method} and \ref{moldat}. Calculated line fluxes for selected frequencies are presented and discussed in Sect.~\ref{result}. The conclusions of this study are drawn in Sect.~\ref{conclusion}.

%__________________________________________________________________

\section{Chemistry}
\label{chemistry}

The chemical network for CO$^+$ has been discussed in the past by many authors \citep[e.g.,][]{StDa95,SaZi04,StDv04,StDv05,StBJ07} and most recently by \citet{BrDV09}. Therefore, we give only a short summary of the most important reactions. 

Under the influence of FUV photons, CO$^+$ is efficiently formed in reactions of C$^+$ with OH. C$^+$ stems from photodissociation and subsequent ionization of CO and C, respectively. OH has high abundances at temperatures above $\approx 300$~K, where large amounts of oxygen are driven into OH due to the dissociation of gaseous water. For low FUV fluxes, other formation routes are the reactions of C$^+$ with CO$_2$ and O$_2$. CO$^+$ rapidly reacts with H$_2$ and H to form HCO$^+$, HOC$^+$ and CO, respectively. Another fast destruction mechanism is the dissociative recombination with electrons. 

\subsection{Parameter study}
\label{parstud}

\begin{figure*}
\centering
\includegraphics[width=17cm]{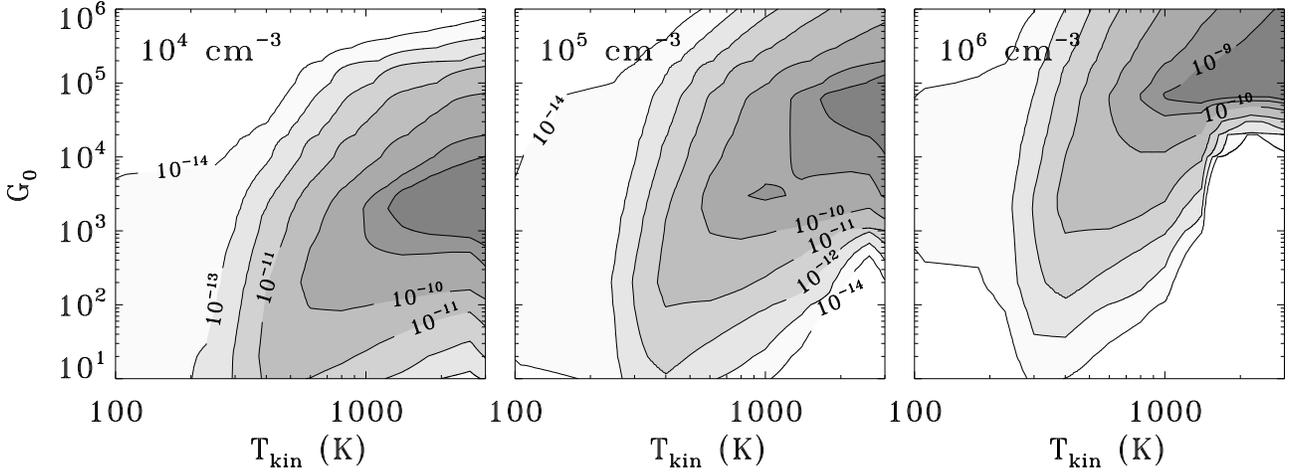}
\caption{Contour levels for the CO$^+$ fractional abundances for hydrogen densities between $10^4$~cm$^{-3}$ and $10^6$~cm$^{-3}$. The assumed optical depth is $\mathrm{A_V} = 0.5$. The contour levels are equally spaced from $10^{-14}$ to $10^{-10}$ and extended by $5\times10^{-10}$ and $10^{-9}$.}
\label{fchem1}
\end{figure*}

The strong dependence on the OH abundance and the fact, that CO$^+$ is quickly destroyed by hydrogen makes its abundance very sensitive to the gas temperature and density. To show this, we have used the chemical grid of \citet{BrDB09} to calculate the fractional abundances of CO$^+$ for different gas temperatures $T_{\mathrm{kin}}$, gas densities $n(\mathrm{H_2})$ and FUV fluxes G$_0$ (in units of the Habing field).  The models of \citet{PeBF09} for Orion Bar showed that CO$^+$ is also sensitive to cosmic rays. We do not further investigate their influence but include a generic cosmic ray ionization rate of $5.6 \times 10^{-17}$\,s$^{-1}$. It should be noted that the chemical equilibrium of CO$^+$ for the densities and temperatures mentioned above is reached very quickly. Our time-dependent chemical models indicate that the CO$^+$ abundance remains constant after $\approx 1000$~years.

Figure~\ref{fchem1} shows the results for gas densities $n(\mathrm{H_2}) = 10^4$--$10^6$~cm$^{-3}$ and an optical depth $\mathrm{A_V} = 0.5$. At this optical depth, CO$^+$ is believed to reach its maximum abundance in dense PDRs \citep{StDa95}. The observed column densities are between $\sim 10^{11}$--$10^{12}$~cm$^{-2}$ (Table~\ref{tobs}). Assuming all CO$^+$ to be produced at $\mathrm{A_V} = 0.5$, these column densities correspond to fractional abundances of $x(\mathrm{CO^+}) \approx 10^{-10}$--$10^{-9}$. According to Fig.~\ref{fchem1}, $x(\mathrm{CO^+}) \approx 10^{-9}$ is produced for $n(\mathrm{H_2}) = 10^4$~cm$^{-3}$ with $T_{\mathrm{kin}} \ga 1000$~K and $G_0 \approx 10^3$--$10^4$; for $n(\mathrm{H_2}) = 10^5$~cm$^{-3}$ with $T_{\mathrm{kin}} \ga 1600$~K and $G_0 \approx 10^4$--$10^5$ and for $n(\mathrm{H_2}) = 10^6$~cm$^{-3}$ with $T_{\mathrm{kin}} \ga 1000$~K and $G_0 \ga 5\times 10^4$. A fractional abundance of $10^{-10}$ can be reached for $n(\mathrm{H_2}) = 10^4$~cm$^{-3}$ with $T_{\mathrm{kin}} \ga 600$~K and $G_0 \approx 10^2$--$10^5$; for $n(\mathrm{H_2}) = 10^5$~cm$^{-3}$ with $T_{\mathrm{kin}} \ga 600$~K and $G_0 \approx 10^3$--$5 \times 10^5$ and for $n(\mathrm{H_2}) = 10^6$~cm$^{-3}$ with $T_{\mathrm{kin}} \ga 600$~K and $G_0 \ga 10^4$. For a typical FUV flux $G_0 \approx 10^4$ in PDRs, abundant CO$^+$ ($x(\mathrm{CO^+}) = 10^{-10}$) can therefore only be expected in regions with gas temperatures $T_{\mathrm{kin}} \ga 600$~K.

\subsection{Comparison with observation}
\label{compobs}

Although the chemical model results presented in Sect.~\ref{parstud} are completely independent of geometry, a simple comparison with observations can be made by assuming the clouds to be large enough in size to provide the observed column densities for $\mathrm{A_V} = 0.5$. The assumption is reasonable since the region needs to be only $1000$--$10\,000$~AU in size for $n(\mathrm{H_2}) = 10^5$~cm$^{-3}$ and $N(\mathrm{CO^+}) = 10^{11}$--$10^{12}$cm$^{-2}$ whereas most PDRs are larger (depending on the viewing angle of course). Table~\ref{tobs} lists CO$^+$ column densities observed toward well known PDRs. Also shown is the ratio of the column density to a hydrogen column density at $\mathrm{A_V} = 0.5$ ($N(\mathrm{H_2}) \approx 2\times 10^{21}$~cm$^{-2}$). This ratio allows direct comparison with the chemical models. It is apparent that the observations are in fair agreement with the model results presented in Fig.~\ref{fchem1}. For example, M$17$SW and Orion Bar have a fractional abundance of $\approx 10^{-9}$ with $N(\mathrm{CO^+}) \approx 10^{12}$~cm$^{-2}$ at $\mathrm{A_V} = 0.5$ in regions with $n(\mathrm{H_2}) \approx 10^5$--$10^6$~cm$^{-3}$ and $G_0 = 5\times 10^4$. Such an abundance is consistent with our models for $T_{\mathrm{kin}} \ga 1000$~K. NGC~$7023$ has a somewhat smaller gas density and $G_0 = 2\times 10^3$. The observed abundance is comparable to our models for $T_{\mathrm{kin}} \ga 600$~K. Mon~R$2$ and G$29.96$--$0.02$ on the other hand are believed to have gas densities $n(\mathrm{H_2}) \approx 10^6$~cm$^{-3}$ and $G_0 \approx 10^5$. The column densities of these dense objects are in agreement with our models for $T_{\mathrm{kin}} \ga 800$~K. 

\subsection{Discussion}
\label{chemdisc}

The parameter study presented above is a simplification of the physics and chemistry in order to study the CO$^+$ abundance for a few effects such as FUV flux, temperature and density. In reality, the conditions for the chemistry is expected to vary from one source to another. For example, \citet{PeBF09} modeled the Orion Bar PDR using the spectral synthesis code Cloudy \citep{FKV98} to derive the physical conditions across the Bar. They assume the cosmic rays to be trapped in the cloud by a tangled magnetic field which eventually yields a highly enhanced cosmic ray ionization rate. This allows them to successfully reproduce the observed CO$^+$ column densities.  However, the simple comparison in the previous section between models and observations shows good agreement if a certain temperature is assumed. The problem of traditional PDR models to produce high CO$^+$ abundances may thus not lie in the chemistry itself but in the uncertainty of the gas temperature at $\mathrm{A_V} \approx 0.5$--$1$ which needs to be $\ga 600$~K. The gas temperature is usually calculated self-consistently in these models by solving the thermal balance. The various PDR codes, however, show differences on the order of a magnitude for $\mathrm{A_V} = 0.5$--$1$, ranging from a few hundred to a few thousand Kelvin \citep{RoAB07}. This uncertainty results eventually in different CO$^+$ abundances. CO$^+$ may thus serve as a tracer for high gas temperatures at low $\mathrm{A_V}$. 

The main uncertainty in the chemical models, on the other hand, are the reaction rates. For example, the rate coefficient for the C$^+$ $+$ OH reaction in our models is $7.7\times 10^{-10}$~cm$^{-3}$~s$^{-1}$,  independent of temperature \citep[UMIST,][]{WoAM06}. The temperature dependent rates published by \citet{DuGM92} and \citet{Troe96} on the other hand, are above $10^{-9}$~cm$^{-3}$~s$^{-1}$. However, the branching ratio is not clear as the C$^+$ $+$ OH $\rightarrow$ CO $+$ H$^+$ reaction is somewhat more exoergic than the CO$^+$ $+$ H channel. It is interesting, that the former reaction, producing H$^+$, may partly recycle OH, because H$^+$ has a high probability of charge transfer with O in neutral gas at the temperatures of interest and the resulting O$^+$ will lead back to OH and H$_2$O as long as the H$_2$/H ratio is not too low \citep[e.g.,][]{StDa95,StJVD06}. \citet{FuGU08} were able to reproduce the observed CO$^+$ abundances with a reaction rate of $2.9\times 10^{-9}$~cm$^{-3}$~s$^{-1}$. Assuming that the abundance roughly scales with this reaction rate, the modeled CO$^+$ abundances may well be $3$--$4$ times higher. 

\begin{table}
\begin{minipage}[t]{\columnwidth}
\caption{CO$^+$ column densities $N$ derived from observations toward PDRs}            
\label{tobs}      
\centering                      
\renewcommand{\thefootnote}{\thempfootnote}  % to avoid a line before footnotes
\begin{tabular}{l c c c c}        
\hline\hline                
Object  & $n(\mathrm{H_2})$ & $G_0$ & $N$  & $N$/$2\times 10^{21}$ \\   
        & [cm$^{-3}$] & [Habing] & [$10^{12}$~cm$^{-2}$] &  [$10^{-9}$] \\  
\hline                        
M$17$SW\footnote{\citet{LaWM93,StSS95}} & $5 \times 10^5$ & $5 \times 10^4$ & $1$--$1.8$ & $0.5$--$0.9$ \\
Orion Bar\footnote{\citet{StSS95,FuRG03,SaZi04}} & $2.5 \times10^5$ & $5 \times 10^4$ & $1$--$2.7$ & $0.5$--$1.35$ \\
S$140$\footnote{\citet{PaMi95,StSS95,SaZi04}} & $10^5$ & $150$ & $0.03$ & $0.015$  \\
NGC~$7023$\footnote{\citet{FuMa97,FuRG03}} & $10^4$--$10^5$ & $2 \times 10^3$ & $0.3$ & $0.15$  \\
Mon~R$2$\footnote{\citet{RiFR03}} & $1.5\times 10^6$ & $4.9 \times 10^5$ & $0.53$ & $0.27$  \\
G$29.96$--$0.02$\footnotemark[\value{footnote}] & $7\times 10^5$ & $1.5 \times 10^5$ & $0.47$ & $0.24$  \\
\hline  
\end{tabular}
\renewcommand{\footnoterule}{}
\end{minipage}
\end{table}
 
A strong support for the argument that the CO$^+$ chemistry may actually be well understood comes also from recent model results by \citet{FuGU08} and \citet{BrDV09} where the geometry of the observed object has been taken into account. \citet{FuGU08} modeled M~$82$ by assuming a plane-parallel cloud illuminated by FUV photons from two sides. With this more realistic view of the interstellar medium in M~$82$, they obtained $\approx 5\times$ larger CO$^+$ column densities in good agreement with the observations and increased the goodness of fit for other observed species. \citet{BrDV09} constructed a two dimensional model of the massive star-forming region AFGL~$2591$ to include FUV irradiated outflow walls. They found that the FUV fluxes and gas temperatures along the walls are high enough to produce large amounts of CO$^+$. Their models were able to reproduce the observed CO$^+$ abundance within a factor of $\approx 2$, whereas the one dimensional spherical models of \citet{StDv04,StDv05} underestimated the CO$^+$ abundance by several orders of magnitude. A reason for this is that a suitable geometry increases the $\mathrm{A_V} \approx 0.5$ area along the line of sight. 
 
X-rays are able to enhance the CO$^+$ abundance to similar values. \citet{StBJ07} and \citet{BrDV09} showed that X-ray fluxes $\mathrm{F_X} \ga 0.1$~erg~s$^{-1}$ lead to fractional abundances $\mathrm{x(CO^+)}\approx 10^{-10}$--$10^{-9}$. However, since most observations of CO$^+$ are believed to trace FUV photons rather than X-rays \citep[e.g.,][]{StBJ07,FuGU08}, the chemistry of CO$^+$ in XDRs is not explored and we refer to the papers mentioned above for further information.  EUV photons with energies above $14$\,eV might be an important source for CO$^+$ at the boundaries of photoionized nebulae. This is not included in the models of \citet{BrDB09} though. EUV photons are quickly absorbed by the gas and the effect on the total CO$^+$ abundance is expected to be small. However, EUV chemistry would only increase the CO$^+$ abundance and thus improve the goodness of fit of chemical models when compared to observations.

Although a careful treatment of the gas temperature and the geometry of the object is crucial to interpret CO$^+$ observations properly, the question remains what the influence of the excitation mechanism is on emission lines. This will be studied in the next section.

%______________________________________________________________

\section{Radiative transfer analysis}
\label{radex}

\subsection{Method}
\label{method}

Since CO$^+$ is destroyed by hydrogen and electrons on a relatively short timescale, the chemical formation and destruction rates need to be considered when calculating the statistical equilibrium \citep[e.g.,][]{vdTBS07}:
\begin{equation}
\frac{d n_i}{dt} = \sum_{i \ne j}^N n_j P_{ji} - n_i \sum_{i \ne j}^N P_{ij} + F_i-D_i = 0 \ , \label{eq:ratn}
\end{equation}
where $N$ is the number of levels considered in the model, $n_i$ [cm$^{-3}$] is the level population of level $i$, $F_i$ and $D_i$ [cm$^{-3}$ s$^{-1}$] are the chemical formation and destruction rates for level $i$, respectively, and $P_{ij}$ [s$^{-1}$] is given by
\begin{equation}
P_{ij} = \left\{
\begin{array}{ll}
A_{ij}+B_{ij} \bar{J_\nu} + C_{ij} & (E_i > E_j) \\
B_{ij} \bar{J_\nu} + C_{ij} & (E_i<E_j) \ . \\
\end{array}
\right.
\end{equation}
$A_{ij}$ and $B_{ij}$ are the Einstein coefficients for spontaneous and induced emission, $C_{ij}$ are the rate coefficients for collisions, $\bar{J_\nu}$ is the specifice intensity integrated over the 
line profile and averaged over all directions and $E_i$ is the energy of level $i$.

Following the results of Sect.~\ref{chemistry}, we assume that the chemical rates are in equilibrium 
($F \equiv \sum_i F_i = \sum_i D_i \equiv D$). 
It is further assumed that CO$^+$ is formed in an excited state, where the levels are 
populated according to a formation temperature $T_{\rm form}$ and that all levels 
are destroyed with equal probability. Hence, the chemical formation and destruction rate for level $i$ are
\begin{eqnarray}
F_i &\equiv& \frac{g_i e^{-E_{i}/k T_{\rm form}}}{\sum_{j=1}^N g_j e^{-E_j/k T_{\rm form}}} F \label{eq:fi}\\
D_i &\equiv& \frac{D}{N} \label{eq:di} \ ,
\end{eqnarray}
where $g_i$ is the statistical weight of level $i$. Equation \ref{eq:ratn} and the equilibrium between formation and destruction ($F = D$) yield
\begin{equation}
0 = \sum_{i \ne j}^N n_j P_{ji} - n_i \sum_{i \ne j}^N P_{ij} + D \left(\frac{g_i e^{-E_{i}/k T_{\rm form}}}{\sum_{j=1}^N g_j e^{-E_j/k T_{\rm form}}} - \frac{1}{N} \right) \ . \label{eq:ratfd1}
\end{equation}
It should be emphasized that the formation temperature is an artifice that allows to describe the effect of the nascent population distribution by a single parameter. The formation temperature should not be conflated with the physical temperature nor with the excitation temperature\footnote{The excitation temperature of a transition in the radiative transfer model is defined by the ratio of the upper level population to the lower one, that is by $x_{i+1}/x_i = g_{i+1}/g_i e^{-h\nu /T_{\mathrm{ex}}}$.}.

For a consistent solution of these equations, the condition $\sum_{i=1}^{N} n_i = n({\rm CO}^+)$ needs to be introduced. It is convenient to scale Eq.~\ref{eq:ratfd1} with $1/n({\rm CO}^+)$ since the equations become independent of $n({\rm CO}^+)$. This scaling does not change the excitation temperature and hence the line intensity obtained from the level populations. It is further justified by the assumption of $F=D$ which ensures that the chemical rates are proportional to $n(\mathrm{CO^+})$ (see also Eq.~\ref{eq:normdestr}). However, the system of equations can now be solved for the fractional abundances of levels ($x_i = n_i / n({\rm CO}^+)$):
\begin{equation}
0 = \sum_{i \ne j}^N x_j P_{ji} - x_i \sum_{i \ne j}^N P_{ij} + \mathcal{D} \left(\frac{g_i e^{-E_{i}/k T_{\rm form}}}{\sum_{j=1}^N g_j e^{-E_j/k T_{\rm form}}} - \frac{1}{N} \right) \ , \label{eq:ratfd2}
\end{equation}
with the normalized destruction rate
\begin{equation}
\mathcal{D} = \frac{D}{n({\rm CO}^+)} = \frac{1}{n({\rm CO}^+)} \sum_i k_i(T_{\rm kin}) n(i) n({\rm CO}^+) = \sum_i k_i(T_{\rm kin})  n(i) \ , \label{eq:normdestr}
\end{equation}
where $n(i)$ is the abundance [cm$^{-3}$] of a molecule $i$ and $k_i(T_{\rm kin})$ is a rate coefficient [cm$^3$ s$^{-3}$] depending on the kinetic temperature. Due to the high abundance of electrons, atomic hydrogen and H$_2$, we can savely take the sum only over these species to calculate the destruction rate:
\begin{equation}
\mathcal{D} \approx 1.5 \times 10^{-9} n({\rm H}_2)+7.5 \times 10^{-10} n({\rm H})+2.0 \times 10^{-7} \left(\frac{T_{\rm kin}}{300 {\rm K}} \right) n(e^-) \ .
\end{equation}
The rate coefficients are taken from the UMIST database for astrochemistry \citep{WoAM06}. According to UMIST, the accuracy of these rates is between $25$\% and $50$\%.
{

To calculate the intensities for a range of rotational transitions, we make use of the publicly available radiative transfer code RADEX \citep{vdTBS07}. RADEX allows to compute quickly the intensities for a grid of defined gas properties in a uniform medium, based on statistical equilibrium calculations. The program includes radiation from background sources and treats optical depth effects with an escape probability method. 

For all calculations, we assume a CO$^+$ column density of $N(\mathrm{CO^+})=10^{12}$~cm$^{-2}$, a line width of $\Delta V = 1$~km~s$^{-1}$ and a $2.73$~K blackbody background. With these values, the lines discussed in the following sections are optically thin. 

\subsection{Molecular data}
\label{moldat}

The energy levels and transition frequencies of CO$^+$ are taken from the JPL database \citep{PiPC98}. The transition probabilities are calculated for a dipole moment $\mu = 2.63$~D \citep{ChBR07}. Since no published excitation rates are available for CO$^+$-H$_2$ collisions, we apply the HCO$^+$-H$_2$ rates of \citet{Flower99}. These rates were calculated for temperatures between $10$~K and $400$~K and rotational levels up to $J=20$. \citet{ScVV05} extrapolated this set of coefficients to include energy levels up to $J=30$ and temperatures up to $2000$~K. We adapt the same rates for the fine-structure levels. In contrast to CO$^+$ ($X~^2\Sigma^+$), HCO$^+$ ($X~^1\Sigma^+$) has a $49$\% higher dipole moment and a $24$\% smaller rotational constant. In addition, the H$_2$ excitation rates are likely to be different due to the nonzero spin of CO$^+$. However, the error may not be bigger than factors $2-3$ \citep[see e.g.,][]{BlaVd91,ScVV05}. For rotational transitions induced by atomic hydrogen, we use the H-CO$^+$ deexcitation rates calculated by \citet{AnBN08} for $J \le 8$. The rates have been extrapolated to levels up to $J=30$ following the method as described by \citet{ScVV05}.  The rate coefficients for electron deexcitation are calculated using Eq.~$2.9$ in \citet{DiFl81}. Since the fine-structure is not treated properly by \citet{DiFl81}, we assume the same rates for the corresponding levels. It should be noted, that \citet{FaT01} have investigated the electron impact excitation of CO$^+$ in more detail. Although they neglect the fine-structure too, it is interesting that the rates for $\Delta N >1$ transitions are significant, something that is not anticipated by \citet{DiFl81}. However, since the formula of \citet{DiFl81} allows to calculate the rates also for high rotational levels, and since electron impact excitation is of minor importance for our purpose, we adopt the treatment of \citet{DiFl81}.

The critical densities ($n_{\mathrm{crit}} = A_{ul}/\sum{K_{ul}}$) of our applied molecular data for $T_{\mathrm{kin}} = 500$~K are presented in Table~\ref{tncrit}. The lines in the table are chosen due to their observability with ground based telescopes. It can be seen from the critical densities that collisions with atomic hydrogen become important for H/H$_2$ ratios of $\ga 1$. Electron excitation will be important for an ionization fraction $x(\mathrm{e^-}) \ga 10^{-3}$.  The critical electron densities are in reasonable agreement (difference is $\la 20$\%) with those published by \citet{FaT01} for $T_{\mathrm{kin}} = 500$~K.

\begin{table}
\caption{Critical densities for $T_{\mathrm{kin}} = 500$~K.}
\label{tncrit}
\centering
\begin{tabular}{l c c c c}
\hline\hline
Species & $N=2\rightarrow 1$ & $N=3\rightarrow 2$ & $N=4\rightarrow 3$ & $N=7\rightarrow 6$ \\
        & $n_{\mathrm{crit}}$ [cm$^{-3}$] & $n_{\mathrm{crit}}$ [cm$^{-3}$] & $n_{\mathrm{crit}}$ [cm$^{-3}$] & $n_{\mathrm{crit}}$ [cm$^{-3}$] \\ 
\hline
H$_2$      & $4.2\times 10^5$ & $1.1\times 10^6$  & $2.1\times 10^6$  & $8.4\times 10^6$    \\
H       & $6.5\times 10^5$ & $1.6\times 10^6$  & $3.0\times 10^6$  & $1.4\times 10^7$    \\
e$^-$   & $3.9\times 10^2$ & $1.4\times 10^3$  & $3.4\times 10^3$  & $1.9\times 10^4$    \\
\hline
\end{tabular}
\end{table}

\subsection{Results and discussion}
\label{result}

\subsubsection{Line fluxes}
\label{fluxes}

The integrated line intensities are calculated for a grid of gas densities ($n(\mathrm{H_2}) = 10^4$--$10^8$~cm$^{-3}$), gas temperatures ($T_{\mathrm{kin}} = 10$--$2000$~K) and formation temperatures ($T_{\mathrm{form}} = 10$--$3000$~K).  Although the line fluxes will also depend on the chemical formation and destruction rates, we treat them as as being well known (Eq.~\ref{eq:normdestr}) and do not vary the chemical rate coefficients, for instance. Also calculated are the fluxes for $\mathcal{D}=0$, that is the radiative transfer models without taking the chemical processes into account. To study the influence of atomic hydrogen and electron excitation, a model was set up with $\mathrm{H/H_2} = 1$ and $x(\mathrm{e^-}) = 10^{-4}$. These are typical values for PDRs with $n(\mathrm{H_2}) \approx 10^5$~cm$^{-3}$ and $G_0 \approx 10^5$. It is found that the results for the $N=2\rightarrow 1$ lines differ $50$\% at most from those where only H$_2$ was considered. For simplicity, atomic hydrogen and electron excitation is thus neglected in the following paragraphs. We have also carried out calculations considering the dust temperature.  Models were run with the dust temperature varying from $20$~K to $1500$~K with a gas density of $10^6$~cm$^{-3}$ and a total H$_2$ column density of $10^{22}$~cm$^{-2}$. The gas temperature was assumed to be the same as the dust temperature. It was found that the influence of dust on the CO$^+$ line fluxes studied here can be neglected under these conditions. Since most observed lines are those with $N=2\rightarrow 1$ and $N=3\rightarrow 2$, we will concentrate on these transitions. 

\begin{figure*}
\centering
\includegraphics[width=17cm]{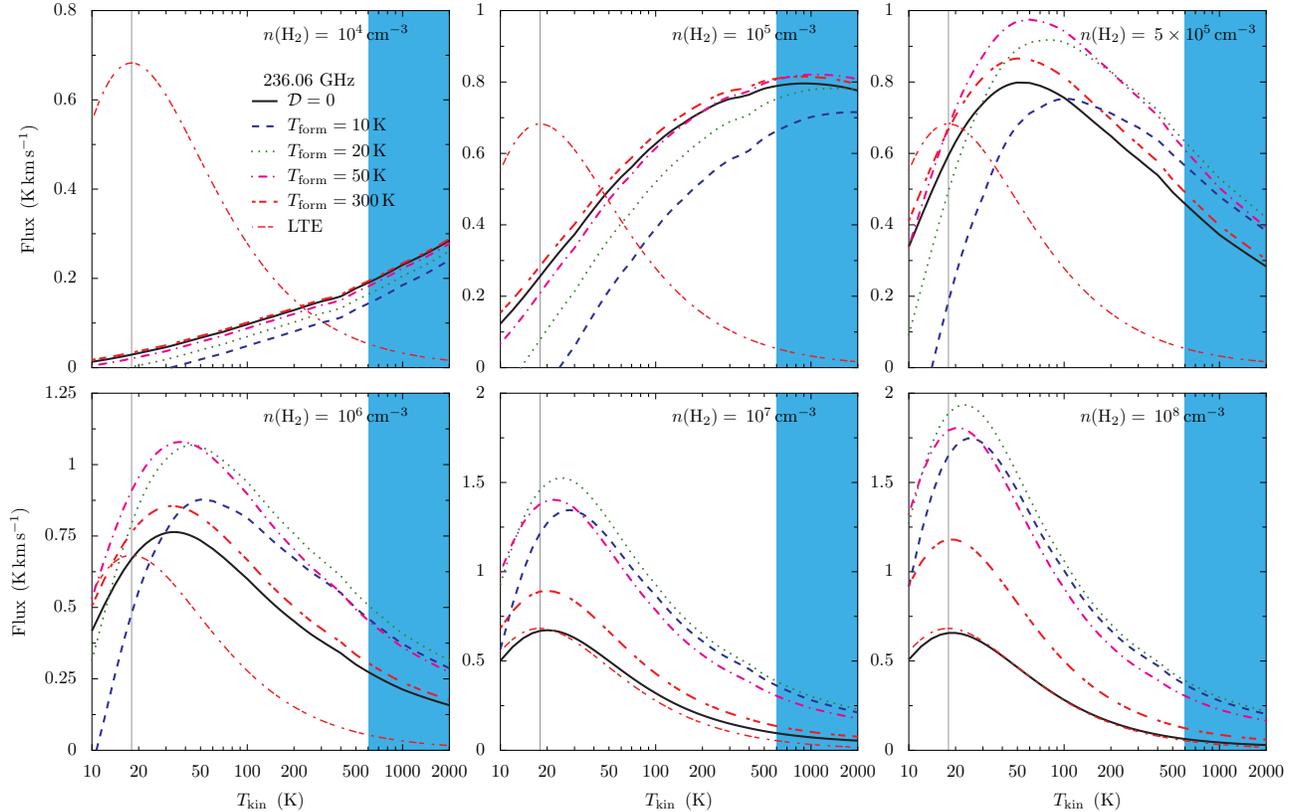}
\caption{Integrated line intensities for the $2\frac{5}{2}\rightarrow 1\frac{3}{2}$ transition at 236~GHz. The solid line corresponds to the radiative transfer models with $\mathcal{D}=0$ (no formation temperature is assumed). The vertical line at $T_{\mathrm{kin}} = 18$~K indicates the position of the peak LTE flux. The shaded region marks the gas temperatures where CO$^+$ is chemically most abundant.}
\label{f236}
\end{figure*}

Figure~\ref{f236} shows the integrated line intensities (K~km~s$^{-1}$) for the $2\frac{5}{2}\rightarrow 1\frac{3}{2}$ transition at $236.06$~GHz for different formation temperatures. Also shown is the flux assuming LTE conditions. The LTE flux was calculated using the analytical expression given in \citet{StBJ07}. As expected, the LTE fluxes are only matched by models with $\mathcal{D}=0$ (no formation temperature assumed) for $n(\mathrm{H_2}) \ga 10^7$~cm$^{-3}$, in other words for densities where collisions dominate and the levels become thermalized. The assumption of a formation temperature $T_{\mathrm{form}} \la 500$~K on the other hand implicates that the levels do not become thermalized (see plot for $n(\mathrm{H_2})=10^8$~cm$^{-3}$ in Fig.~\ref{f236}). 

The results for the different formation temperatures vary only $\approx 20$--$30$\% for low densities ($n(\mathrm{H_2}) \la 5\times 10^5$~cm$^{-3}$) and temperatures where CO$^+$ has high abundances ($T_{\mathrm{kin}} \ga 600$~K; shaded region in the figure). At higher densities, the models including the chemical formation and destruction rates show factors $2$--$7$ higher line fluxes. Models with formation temperatures between $500$~K and $1000$~K have similar fluxes like those with $\mathcal{D}=0$. This is also true for models with formation temperatures exceeding $1000$~K for $n(\mathrm{H_2}) \la 10^7$~cm$^{-3}$. At higher densities, the upper levels of the $N=2\rightarrow 1$ lines are not or only weakly populated for formation temperatures exceeding $1000$~K.

\begin{figure*}
\centering
\includegraphics[width=17cm]{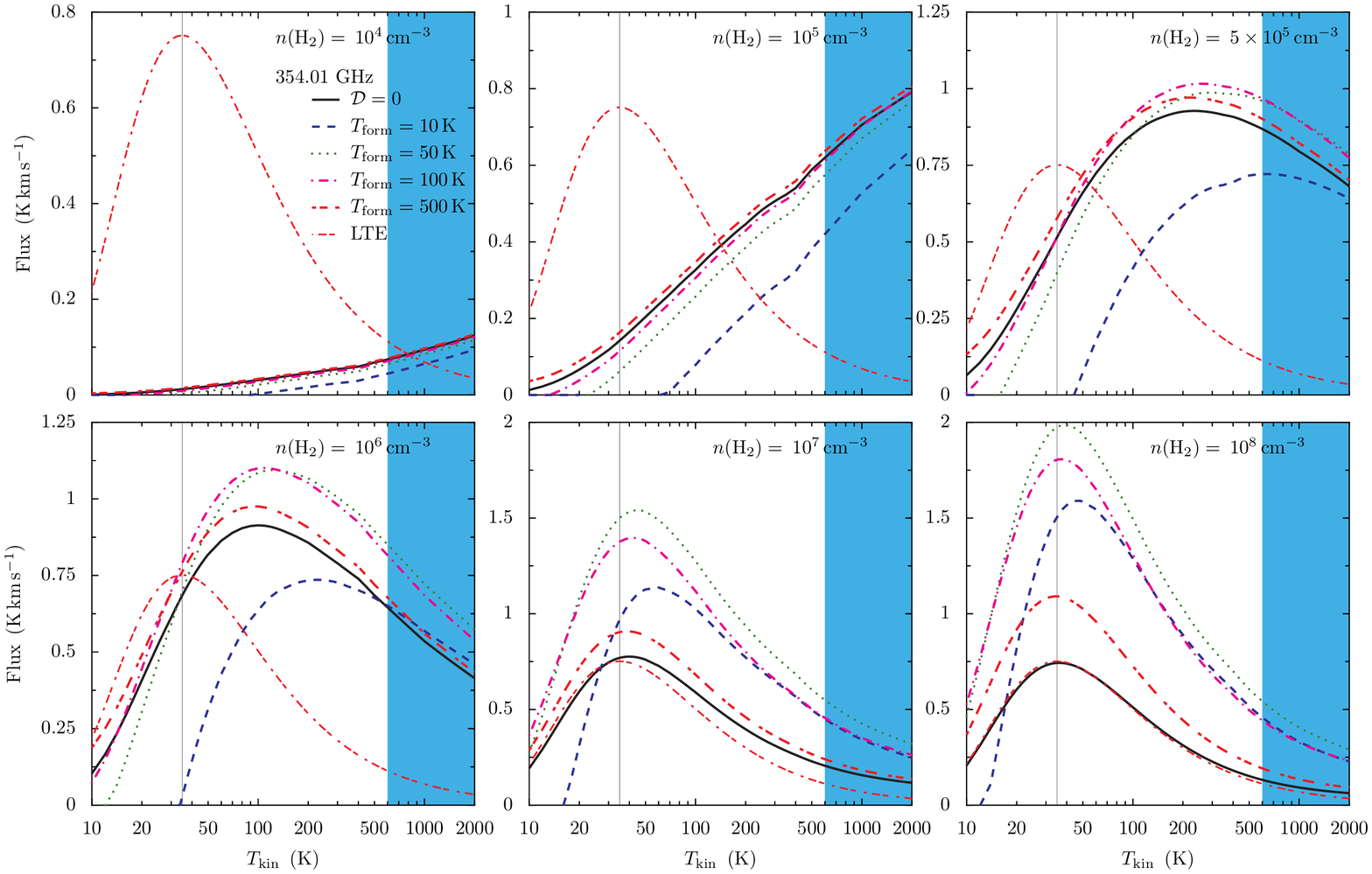}
\caption{Integrated line intensities for the $3\frac{7}{2}\rightarrow 2\frac{5}{2}$ transition at 354~GHz. The solid line corresponds to the radiative transfer models with $\mathcal{D}=0$ (no formation temperature is assumed). The vertical line at $T_{\mathrm{kin}} = 35$~K indicates the position of the peak LTE flux. The shaded region marks the gas temperatures where CO$^+$ is chemically most abundant.}
\label{f354}
\end{figure*}

Figure~\ref{f354} shows the results for the $3\frac{7}{2}\rightarrow 2\frac{5}{2}$ transition at $354.01$~GHz. The fluxes for the different formation temperatures vary only $\approx 20$--$40$\% for densities $\la 10^6$~cm$^{-3}$. At higher densities, the fluxes can be up to $3$ times higher, depending on the formation temperature. Models with formation temperatures $\ga 500$~K have similar line fluxes as models where $\mathcal{D}=0$. It can be concluded that the influence of an anomalous excitation temperature on the line flux is small for the CO$^+$ transitions and regions observed so far ($n(\mathrm{H_2}) \la 10^6$~cm$^{-3}$).

Bigger differences may be expected for transitions with upper energy levels comparable to the gas temperature where CO$^+$ is most abundant. The high spectral resolution instrument HIFI on board the Herschel Space Observatory covers the frequency range $480$--$1910$~GHz and thus the upper CO$^+$ levels $N_{\mathrm{up}}=5$--$16$ with upper energy levels between $85$~K and $769$~K. Figure~\ref{N86} shows the line fluxes for transitions $N \rightarrow N-1$ of the first $16$ upper levels for $n(\mathrm{H_2}) = 10^8$~cm$^{-3}$ and $T_{\mathrm{kin}} = 600$~K. For the sake of clarity, only the stronger one of the two fine-structure lines are presented. The density of $10^8$~cm$^{-3}$ represents the critical density of the higher $N_{\mathrm{up}}$-levels. It can be seen that the various formation temperatures result in significantly different line fluxes. For example, transitions with $N_{\mathrm{up}} \ge 14$ have an order of magnitude lower line fluxes when the formation temperature is much lower than the corresponding upper energy level ($T_{\mathrm{form}} \approx 10$~K). Further calculations for $n(\mathrm{H_2}) = 10^6$~cm$^{-3}$ show that this is already the case for $N_{\mathrm{up}} \ge 9$. Calculations for $n(\mathrm{H_2}) = 10^4$~cm$^{-3}$, on the other hand, indicate that transitions with $N_{\mathrm{up}} \ge 7$ have line fluxes below $10$~mK and may therefore not be observable for $N(\mathrm{CO^+}) = 10^{12}$~cm$^{-3}$. Interestingly, \citet{CeCW97} detected far--infrared CO$^+$  lines up to $N=21$ towards the low-mass star-forming region IRAS $16293$--$2422$. Clearly, the formation temperature of these CO$^+$ levels is either high or there is no nascent excitation due to direct ionization of CO. This might indicate the presence of X-rays, enhanced cosmic rays or EUV photons. Nevertheless, dense PDRs and protoplanetary disk atmospheres with high gas densities ($\ga 10^6$~cm$^{-3}$) and high temperatures ($\ga 600$~K) are ideal testbeds to study the excitation of CO$^+$ in the terahertz wavelength regime with Herschel. 

Similar might be true for other molecular ions that will be observed with Herschel. CH$^+$, for example, is also more likely to be destroyed than excited by hydrogen \citep{Black98}. The chemistry for CH$^+$, however, is more complex. The main source of CH$^+$ is the endoergic reaction C$^+$ $+$ H$_2$ $\rightarrow$ CH$^+$ $+$ H. Besides high gas temperatures, vibrationally excited hydrogen could carry a significant part of the energy needed to activate the reaction \citep{StDa95}. This introduces another uncertainty to chemical models. In addition, CH$^+$ is found to be sensitive to the thermal emission by dust \citep{Black98}. Nevertheless, care should be taken when analyzing observations of such reactive ions. They do not necessarily reflect the column density and hence the molecular abundance when they are excited upon formation.
%However, other reactive molecular ions such as OH$^+$, H$_2$O$^+$ and NH$^+$ are likely to show a similar excitation behavior and relatively simple chemistry like CO$^+$.

\begin{figure}
\resizebox{\hsize}{!}{\includegraphics{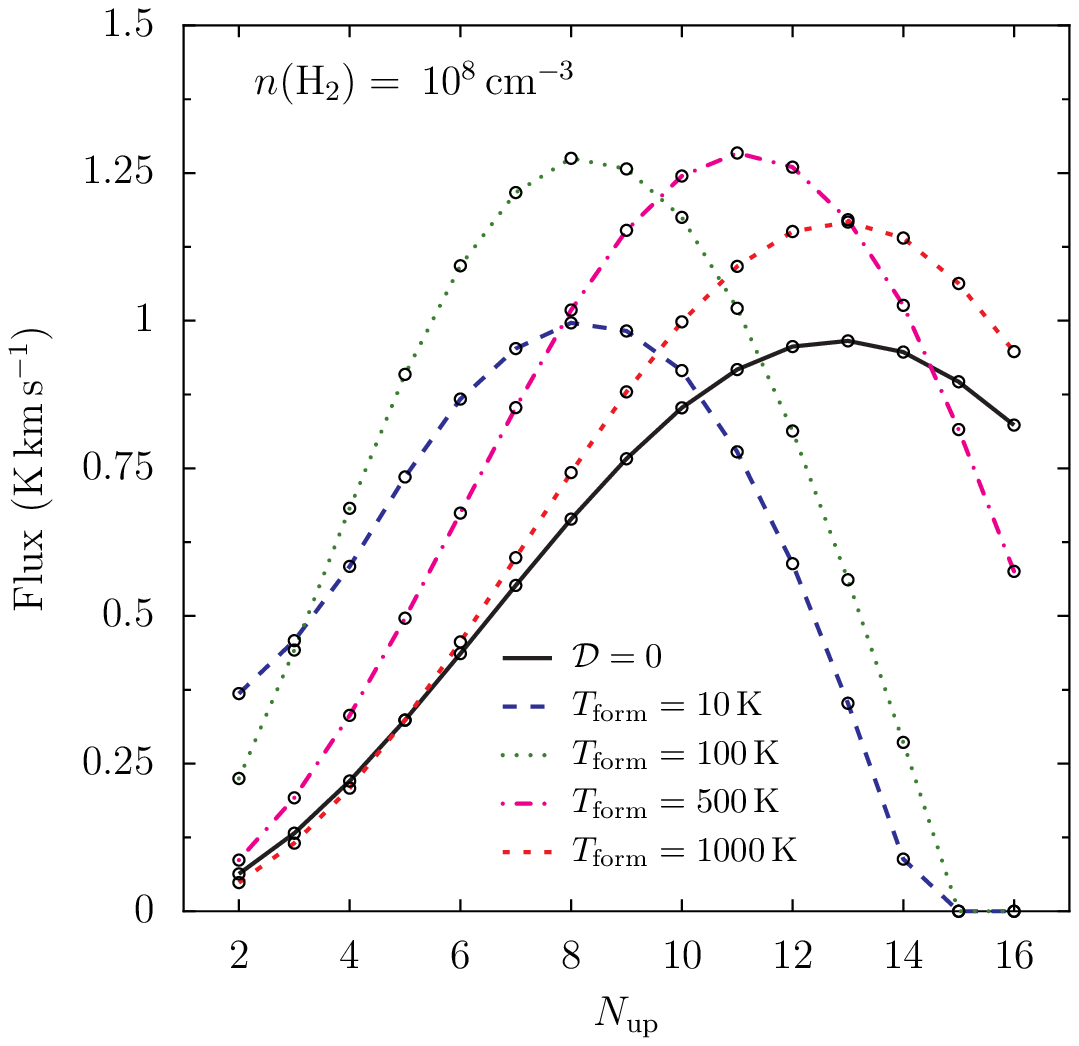}}
\caption{Line fluxes for different formation temperatures are given as a function of $N_{\mathrm{up}}$ for the transitions $N_{\mathrm{up}} \rightarrow N_{\mathrm{up}}-1$.  Of the two fine-structure lines, only the stronger one is presented. The solid line corresponds to the models where $\mathcal{D}=0$. The following formation temperatures are considered: $T_{\mathrm{form}}=10$~K (dashed line), $T_{\mathrm{form}}=100$~K (dotted line), $T_{\mathrm{form}}=500$~K (dash-dotted line), $T_{\mathrm{form}}=1000$~K (short-dashed line). The gas temperature is assumed to be $600$~K and the CO$^+$ column density $10^{12}$~cm$^{-2}$.}
\label{N86}
\end{figure}

\subsubsection{Excitation temperatures}
\label{extemp}

\begin{figure}
\resizebox{\hsize}{!}{\includegraphics[]{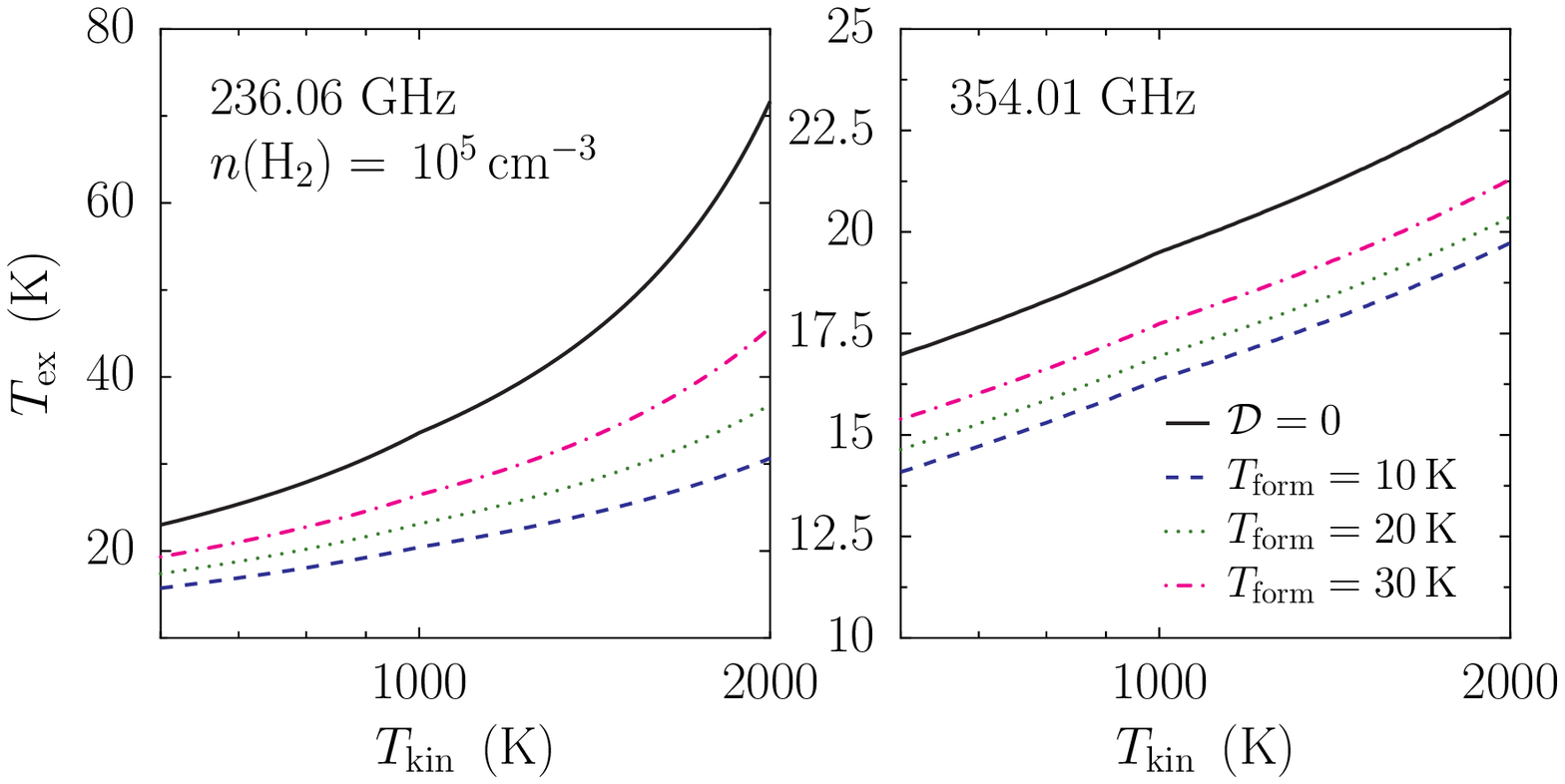}}
\caption{Calculated excitation temperatures for the $2\frac{5}{2}\rightarrow 1\frac{3}{2}$ ($236$~GHz) and $3\frac{7}{2}\rightarrow 2\frac{5}{2}$ ($354$~GHz) transitions for $n(\mathrm{H_2}) = 10^5$~cm$^{-3}$. The solid line corresponds to the radiative transfer models where no formation temperature is assumed ($\mathcal{D}=0$).}
\label{ftex}
\end{figure}

It is interesting to see that the calculated line fluxes for gas densities and gas temperatures of typical dense PDRs ($n(\mathrm{H})$ of a few $\times 10^5$~cm$^{-3}$ and $T_{\mathrm{kin}} \ga 600$~K) are comparable to the maximal LTE line fluxes (see LTE fluxes at $T_{\mathrm{kin}} = 18$~K and $T_{\mathrm{ex}} = 35$~K in Fig.~\ref{f236} and Fig.~\ref{f354}, respectively). For these conditions, the assumption of a low excitation temperature therefore yields approximately the same flux as detailed non--LTE calculations.

It has already been mentioned, that the observed CO$^+$ lines show rather low ($T_{\mathrm{ex}} \approx 10$~K) excitation temperatures \citep{LaWM93,FuRG03}. This agrees well with our models. The calculated non--LTE excitation temperatures for a CO$^+$ column density of $10^{12}$~cm$^{-2}$ and $n(\mathrm{H_2}) = 10^5$~cm$^{-3}$} are between $\approx 15$~K and $\approx 70$~K (Fig.~\ref{ftex}) for gas temperatures between $600$~K and $2000$~K. The excitation temperatures for the $N=3\rightarrow 2$ line are equally low ($T_{\mathrm{ex}} \approx 10$--$25$~K). They are clearly below the gas temperature and thus far from LTE. 

The assumption of a formation temperature with $T_{\mathrm{form}} = 10$~K leads to an excitation temperature that matches the observed one best. This suggests that the CO$^+$ lines observed by \citet{LaWM93} and \citet{FuRG03} were excited upon formation with an excitation energy corresponding to a low formation temperature.

\subsubsection{Line ratios}
\label{ratios}

\begin{figure}
\resizebox{\hsize}{!}{\includegraphics{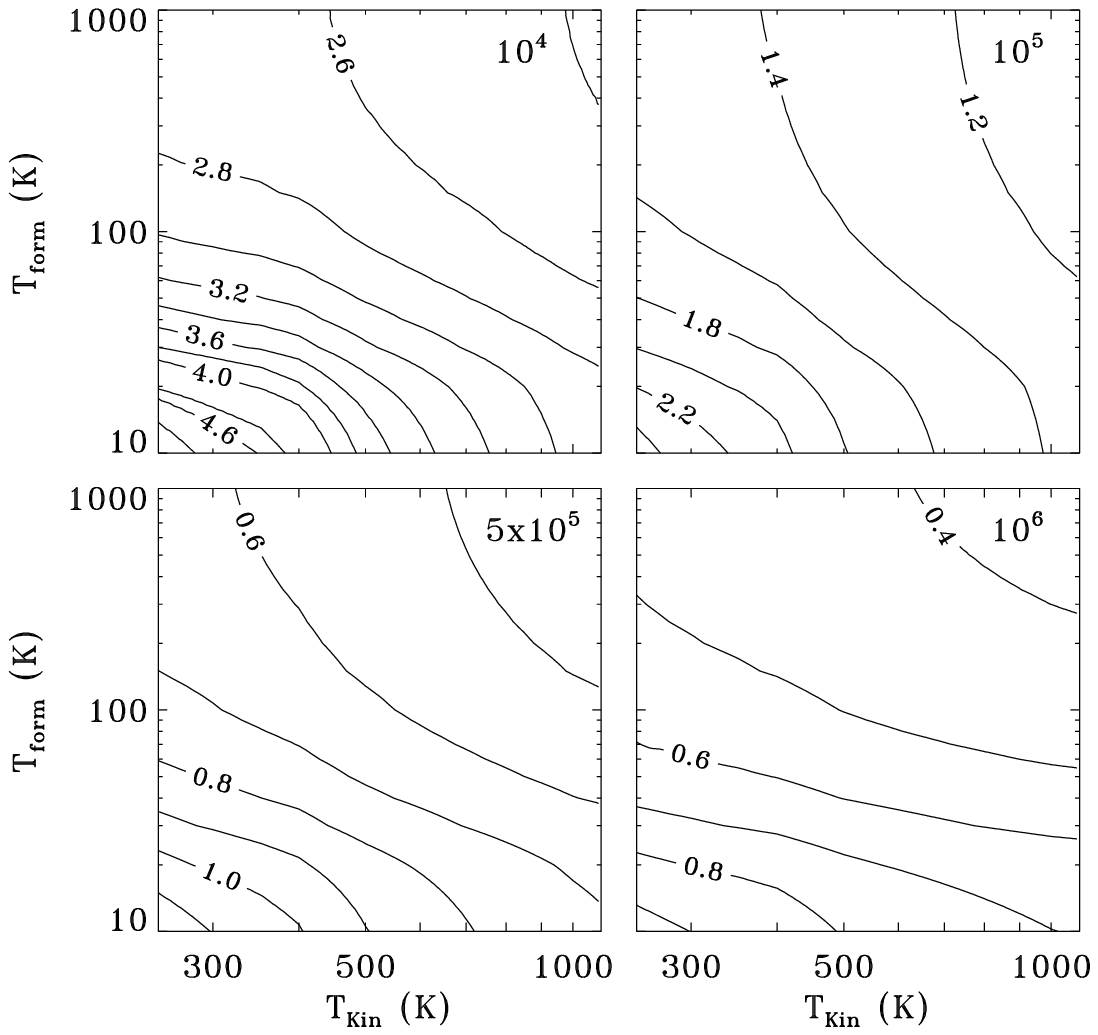}}
\caption{The $J=2\frac{5}{2}\rightarrow 1\frac{3}{2}/3\frac{7}{2}\rightarrow 2\frac{5}{2}$ ($236/354$~GHz) line ratios are shown for the densities $10^4$, $10^5$, $5\times 10^5$ and $10^6$~cm$^{-3}$.}
\label{c23}
\end{figure}

\begin{figure}
\resizebox{\hsize}{!}{\includegraphics{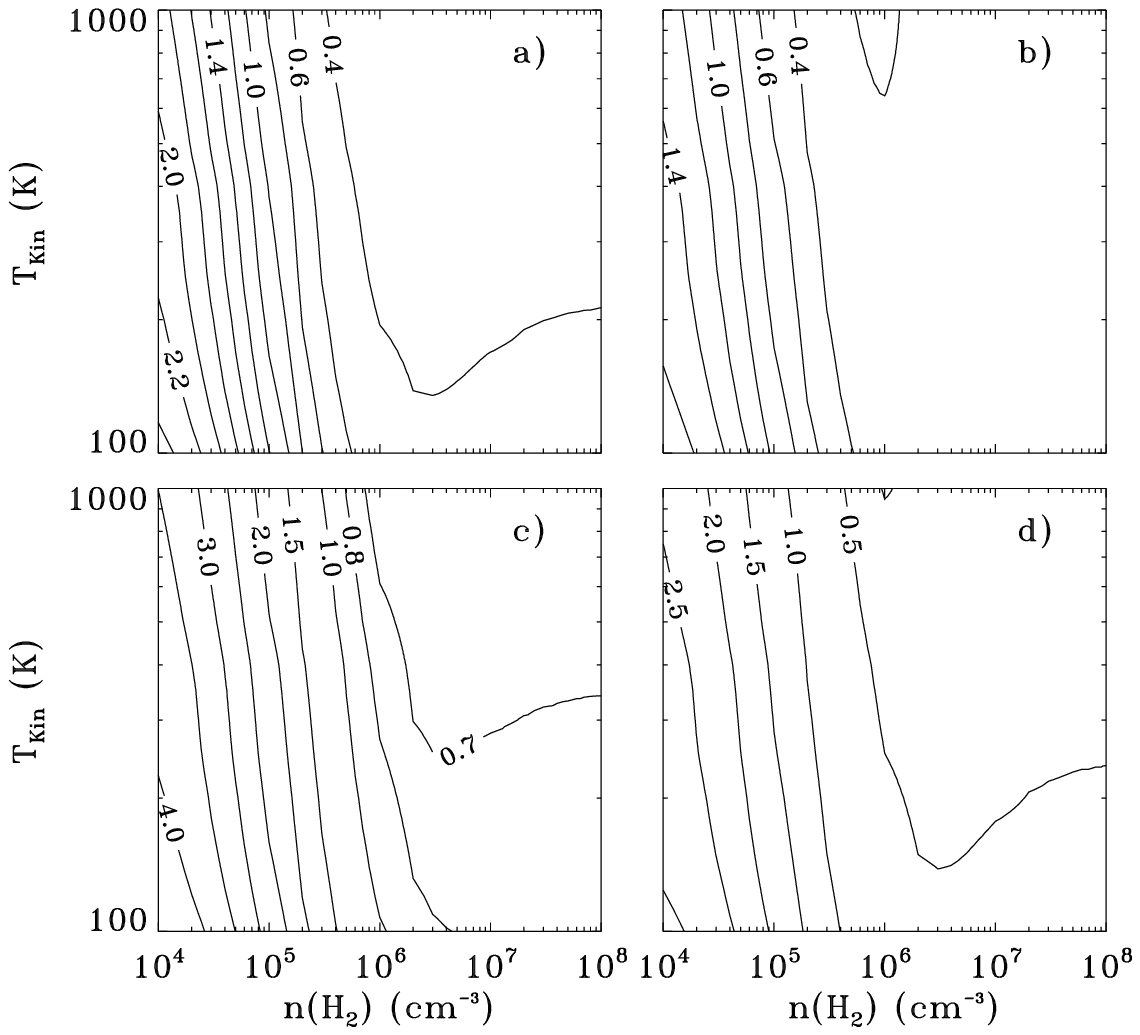}}
\caption{The plots from top left to bottom right show the flux ratios for a) $J=2\frac{3}{2}\rightarrow 1\frac{1}{2}/3\frac{5}{2}\rightarrow 2\frac{3}{2}$ ($235/353$~GHz) b)  $J=2\frac{3}{2}\rightarrow 1\frac{1}{2}/3\frac{7}{2}\rightarrow 2\frac{5}{2}$ ($235/354$~GHz) c) $J=2\frac{5}{2}\rightarrow 1\frac{3}{2}/3\frac{5}{2}\rightarrow 2\frac{3}{2}$ ($236/353$~GHz) and d) $J=2\frac{5}{2}\rightarrow 1\frac{3}{2}/3\frac{7}{2}\rightarrow 2\frac{5}{2}$ ($236/354$~GHz) as a function of the gas temperature and H$_2$ density.}
\label{c023}
\end{figure}

Line ratios are useful to estimate physical gas conditions. Figure~\ref{c23} shows the contour lines for different densities for the $J=2\frac{5}{2}\rightarrow 1\frac{3}{2}/3\frac{7}{2}\rightarrow 2\frac{5}{2}$ flux ratio as a function of the gas temperature and formation temperature. The contour lines for the flux ratio depending only on gas temperature and density ($\mathcal{D}=0$) are given in Fig.~\ref{c023}. It is seen in general that the dependence on the formation temperature is strongest for low gas temperatures. The results for the models without taking the chemical formation into account are similar to those with high formation temperatures. 

Comparison of Figs.~\ref{c23} and \ref{c023} reveals that ratios $J=2\frac{5}{2}\rightarrow 1\frac{3}{2}/3\frac{7}{2}\rightarrow 2\frac{5}{2} \ga 3$ are only possible for $n(\mathrm{H_2}) \approx 10^4$~cm$^{-3}$, $T_{\mathrm{kin}} \le 1000$~K and formation temperatures between $10$~K and $100$~K. Such high ratios are thus indicative for an anomalous excitation mechanism with low formation temperatures.

The flux ratios for the two $N=2\rightarrow 1$ and $N=3\rightarrow 2$ fine structure lines are $\approx 0.5$ and $\approx 0.6$--$0.7$, respectively. Their dependence on gas temperature and density is weak as expected in the optically thin limit. In addition, the ratios differ at most 20\% for various formation temperatures and are thus not suitable to estimate the gas conditions or formation temperature. 

\subsubsection{Comparison with observations}
\label{comparison}

In general, the derived column densities from the observed line fluxes are consistent with our model results presented in the previous section. For example, \citet{FuRG03} obtained a column density of $N(\mathrm{CO^+}) = 10^{12}$~cm$^{-2}$ for an observed flux of $0.63$~K~km~s$^{-1}$ toward Orion Bar. This flux is comparable to the radiative transfer models for $n(\mathrm{H_2}) \approx 10^5$~cm$^{-3}$, $T_{\mathrm{kin}} \approx 600$~K and $T_{\mathrm{form}} \approx 10$~K. 

However, when comparing observations with the homogeneous radiative transfer models described in the previous section, four parameters need to be fitted: the column density, the formation temperature and the gas density and temperature. Therefore, detections of at least four different lines are necessary. Unfortunately, more than two observed CO$^+$ transitions are reported only for Orion Bar \citep{HoJv95} and M$17$SW \citep{LaWM93}. Nevertheless, the CO$^+$ fluxes observed toward these two regions are modeled using a $\chi^2$ test as described in \citet{HoJv95}. The set of parameters which minimizes $\chi^2$ is then considered as best fit to the data. Since the critical densities are similar for the lines that are accessible by ground based telescopes, CO$^+$ is not the ideal molecule to trace the gas density. The density is thus adapted as described in the following paragraphs.

The molecular cloud M$17$SW lies to the southwest within the \ion{H}{II} region of the Omega Nebula. The PDR is separating the ionized gas from the molecular cloud. \citet{MeHT92} found that the PDR consists of clumps with $n(\mathrm{H_2})=2.5\times 10^5$~cm$^{-3}$ and $T \approx 1000$~K and a lower density core surrounding the clumps with $n(\mathrm{H_2})=1500$~cm$^{-3}$ and $T \approx 200$~K. The same authors estimated a FUV field of $G_0 \approx 5\times 10^4$. In a homogeneous model, the density was found to be $1.5\times 10^4$~cm$^{-3}$ and the temperature $300$~K. Thus, for the radiative transfer models, a density of $n(\mathrm{H_2})=1.5\times 10^4$~cm$^{-3}$ is assumed with a H/H$_2$ ratio of $1$ and an electron fraction $x(\mathrm{e^-})=10^{-4}$. The error in the observations for the $\chi^2$ test is taken to be $30$\% of the observed flux for the $J=2\frac{5}{2}\rightarrow 1\frac{3}{2}$ lines and $50$\% for the others. The fits with $\chi^2 \le 1$ have column densities $N(\mathrm{CO^+})=1.5\times 10^{12}$--$4.5\times 10^{12}$~cm$^{-2}$ and gas temperatures $\ga 100$~K in good agreement with the values presented in Table~\ref{tobs}. Unfortunately, no conclusions can be drawn regarding the formation temperature. However, the flux ratio of the observed $J=2\frac{5}{2}\rightarrow 1\frac{3}{2}$ and $J=3\frac{5}{2}\rightarrow 2\frac{3}{2}$ lines including the error is $\approx 5$--$8$ \citep{LaWM93}. Further calculations show that this indicates gas temperatures $\la 800$~K and a formation temperature between $10$--$70$~K.

The Orion Bar is a dense molecular ridge within the Orion Molecular Cloud illuminated by the Trapezium stars. From millimeter and submillimeter observations, \citet{HoJv95} find that the Orion Bar is best described by a clumpy medium where $\approx 10$\% of the material may be in clumps with $n(\mathrm{H_2})\approx 10^6$~cm$^{-3}$ and $\approx 90$\% in a homogeneous interclump medium with $n(\mathrm{H_2})\approx 3\times 10^4$~cm$^{-3}$. The FUV field is estimated to be $G_0 = 3\times 10^4$ \citep{JaSH95}. To model the observed CO$^+$ emission, we use a weighted average density of $1.3\times 10^5$~cm$^{-3}$ with a H/H$_2$ ratio of $1$ and $x(\mathrm{e^-})=10^{-4}$. The line widths and fluxes for the two $N=2\rightarrow 1$ and $N=3\rightarrow 2$ transitions, respectively, are taken from the observations of \citet{HoJv95}. The error for the $\chi^2$ test is taken to be $30$\% for the $N=2\rightarrow 1$ transitions and $50$\% for the $N=3\rightarrow 2$ lines since these profiles are dominated by instrumental broadening according to \citet{HoJv95}. Best fit models ($\chi^2 \le 0.8$) show column densities $N(\mathrm{CO^+})=5.5\times 10^{11}$--$7\times 10^{11}$~cm$^{-2}$ and gas temperatures $T_{\mathrm{kin}} \ga 700$~K. The models, however, are not conclusive regarding the formation temperature. Nevertheless, the fits are in good agreement with the chemical model results in Sect.~\ref{chemistry} and with previously reported column densities (Table~\ref{tobs}). 

The low observed excitation temperatures and the high $J=2\frac{5}{2}\rightarrow 1\frac{3}{2}/3\frac{5}{2}\rightarrow 2\frac{3}{2}$ flux ratio \citep{LaWM93,FuRG03} indicate that CO$^+$ may be formed rotationally excited with a low formation temperature ($T_{\mathrm{form}} \la 70$~K). Due to the assumptions made in the models (HCO$^+$-H$_2$ rates, homogeneous medium) we consider this as indication rather than proof though. One should also bear in mind that the chemistry and excitation conditions may vary from source to source. If CO$^+$ is mainly produced by ionizations of CO, no nascent excitation effects will occur as stated in Sect.~\ref{intro}. This is the case by the presence of X-rays \citep{StDv05} or cosmic rays \citep{PeBF09}. UV photons will have this effect only at the boundary of a PDR and is therefore expected to be small. Certainly, more observations of different transitions are needed toward PDRs to further investigate this problem. In addition, multi-dimensional radiative transfer models in combination with chemical models are necessary to deal with the complex geometry of the emitting objects (Bruderer et al., in preparation).

%______________________________________________________________

\section{Conclusion}
\label{conclusion}
A grid of time-dependent chemical models has been calculated to constrain the physical parameters where CO$^+$ is most abundant. Line fluxes have been computed for $N(\mathrm{CO^+}) = 10^{12}$~cm$^{-2}$ by the use of a non--LTE radiative transfer code to study the effects of an anomalous excitation mechanism. The following list summarizes the main conclusions to be drawn from this study:

\begin{list}{}{}
\item $1.$ High fractional abundances ($x(\mathrm{CO^+}) \ga 10^{-10}$ for $\mathrm{A_V}=0.5$) are only reached for gas temperatures $\ga 600$~K. A simple comparison between chemical models and observations shows good agreement if a certain gas temperature is assumed. This suggests that the CO$^+$ chemistry is well understood and that the molecular ion serves as a tracer for hot gas in regions with $\mathrm{A_V}\approx 0.5$ (Sect.~\ref{parstud}). 

\item $2.$ The model results in Sect.~\ref{fluxes} indicate that the influence of an anomalous excitation mechanism is small on $N=2\rightarrow 1$ and $N=3\rightarrow 2$ line fluxes in regions where CO$^+$ has been observed thus far (PDRs with  $n(\mathrm{H_2})\approx 10^5$~cm$^{-3}$). The calculated line fluxes are in good agreement with observations and confirm the chemical model results (Sect.~\ref{comparison}).  

\item $3.$ Formation temperatures in the range of $10$--$1000$~K show significantly different results for transitions with $N_{\mathrm{up}} \ge 9$ for $n(\mathrm{H_2}) = 10^6$~cm$^{-3}$ and $N_{\mathrm{up}} \ge 14$ for $n(\mathrm{H_2}) = 10^8$~cm$^{-3}$ (Sect.~\ref{fluxes}). The Herschel Space Observatory covers the upper CO$^+$ levels $N_{\mathrm{up}}=5$--$16$ and is thus ideally suited to study the rotational excitation of CO$^+$ and related molecular ions.

\item $4.$ Rotational CO$^+$ levels with temperatures much higher than the formation temperature ($E_{\mathrm{up}}/k \gg T_{\mathrm{form}}$) will be scarcely populated when excited upon formation. Transitions from these levels will therefore have low line fluxes and may not be detectable (Sect.~\ref{fluxes}).

\item $5.$ The low excitation temperatures which are observed for CO$^+$ are consistent with the radiative transfer models (Sect.~\ref{extemp}). Comparison with observations suggest formation temperatures of only $\approx 10$~K.

\item $6.$ The $N=3 \rightarrow 2$/$2 \rightarrow 1$ line ratios are found to be enhanced in Sect.~\ref{ratios} for low formation temperatures ($T_{\mathrm{form}} \la 70$~K).

\end{list}

%______________________________________________________________

\begin{acknowledgements}
The authors are grateful to John Black for the valuable comments and suggestions that helped to improve this paper. We further thank Steven Doty, Arnold Benz and Susanne Wampfler for useful discussions. The work was supported by the Swiss National Science Foundation grant $200020$--$113556$.
\end{acknowledgements}

\bibliographystyle{aa}
\bibliography{/Users/p-like/Documents/Lesmots/pst}

\end{document}